\documentclass[aps,reprint,floatfix,citeautoscript,longbibliography,superscriptaddress]{revtex4-1}
\usepackage{graphicx}
\usepackage{bm,amsmath,amssymb,mathrsfs,dcolumn}
\usepackage{txfonts}
\usepackage{gensymb}
\usepackage{epstopdf}
\usepackage{epsfig}
\usepackage[usenames]{color}
\usepackage{subfigure}
\usepackage{ulem}
\usepackage{float}
\usepackage{soul,xcolor}
\usepackage{cancel}
\usepackage{hyperref}
\usepackage[marginal]{footmisc}
\hypersetup{colorlinks=true, linkcolor=blue, citecolor=blue, urlcolor=blue}
\setstcolor{red}

\setcitestyle{open={[},close={]},citesep={,\!},numbers}
\begin{document}
\preprint{APS/123-QED}

\title{Digital Twins of Networked Dynamical Systems: Machine Learning Prediction of Synchronization Transitions in Coupled Chaotic Oscillators}

\author{Huawei Fan}
\email{huawei.fan@xupt.edu.cn}
\affiliation{School of Science, Xi'an University of Posts and Telecommunications, Xi'an 710121, China}

\author{Yao Du}
\affiliation{School of Physics and Information Technology, Shaanxi Normal University, Xi'an 710062, China}

\author{Meng-Xia Bi}
\affiliation{School of Science, Xi'an University of Posts and Telecommunications, Xi'an 710121, China}

\author{Yafeng Wang}
\affiliation{Nonlinear Research Institute, Baoji University of Arts and Sciences, Baoji 721016, China}

\author{Xingang Wang}
\affiliation{School of Physics and Information Technology, Shaanxi Normal University, Xi'an 710062, China}

\begin{abstract}
Collective behaviors such as synchronization transitions are emergent properties of coupled oscillator systems, and how to reproduce them from data for large-scale systems remains an open problem. Here we propose a model-free coupled reservoir computing (CRC) architecture for constructing digital twins of coupled oscillator systems. The CRC decomposes the twin into three modules that mimic the nodal dynamics, the coupling function, and the coupling network, respectively, so that the modules can be reassembled for different coupled systems. Importantly, it is found that the trained CRC reproduces not only the collective states, including complete synchronization, phase synchronization, and chimera states, but also the synchronization transitions as the coupling strength varies, and the predicted collective behaviors are practically identical to those obtained by direct simulations of the original system. Moreover, the CRC is trained only on the time series of an isolated oscillator and of a pair of coupled oscillators, and the trained modules can be reused for different complex networks and for non-identical oscillators. Our findings establish the CRC architecture as a promising platform for digital twins of complex systems.
\end{abstract}

\maketitle

Digital twin, defined as a virtual representation that mirrors the states of a physical system, has attracted continuous interest across multiple disciplines over the past decades~\cite{TQ:2019,RSK:2020}. By integrating observational data with machine learning models, a digital twin enables real-time prediction, monitoring, and control of the underlying system~\cite{KWGH:2023,L:2024}, and has been widely applied in areas such as smart manufacturing~\cite{TIES:2011}, healthcare~\cite{BSV:2018}, and earth system science~\cite{LFRS:2023}. The core of such a twin, however, is a data-driven model that faithfully reproduces the system dynamics. Building digital twins of complex systems, however, remains a challenge, particularly the collective dynamical behaviors of coupled systems.

The fidelity of such a twin is determined by the properties it reproduces, and can be graded into the following four levels. (1) The evolution behavior: predicting the system state at the next time step, as in time series prediction~\cite{JH:2004,PHGL:2018,SCHA:2022,FJZW:2020}. (2) The statistical behavior: reproducing, e.g., the attractor, the Lyapunov exponents, and the probability distributions~\cite{PLHG:2017,PCGP:2021}. (3) The parameter-dependent behavior: reproducing the variation of the system dynamics with respect to the parameters~\cite{KFGL:2021,KLNP:2021,KFGL2:2021,FKLW:2021}, e.g., bifurcation diagrams and phase transitions. (4) The regulation and response behavior: reproducing the system response to external driving and control~\cite{KWGH:2023,TYHN:2019,WYGZ:2019,GZLF:2021}. In the ideal limit, the digital twin approaches the physical system with unlimited fidelity. It is worth noting that the prediction of synchronization transitions requires the twin to reproduce not only the parameter-dependent behavior but also the regulation and response behavior. More specifically, the synchronization transition, occurring as the coupling strength varies, is a parameter-dependent collective behavior, whereas the synchronized motion itself is the collective response of the oscillators to the mutual regulation conveyed by the couplings~\cite{BKOV:2002,ADKM:2008}.

Whereas machine learning has achieved all four levels of fidelity for individual chaotic systems, its extension to large-scale coupled oscillator systems has so far been limited to the first level, that is, the prediction of the evolution behavior~\cite{PHGL:2018,SCHA:2022}. Reaching the higher levels of fidelity for a coupled system is a fundamentally different task~\cite{FKLW:2021}. Synchronization is an emergent behavior: it is not contained in the dynamics of any single oscillator but arises from the interactions among them~\cite{BKOV:2002,ADKM:2008}. As the coupling strength increases, a network of coupled oscillators may undergo one or more synchronization transitions, giving rise to a variety of collective states~\cite{PC:1998,RPK:1996,AS:2004,PSHM:2014}. Reproducing the collective dynamics therefore calls for a data-driven model that incorporates the nodal dynamics, the coupling function, and the coupling network. It remains an open problem to predict the synchronization transitions of large-scale coupled systems.

\begin{figure}[tbp]
\begin{center}
\includegraphics[width=0.8\linewidth]{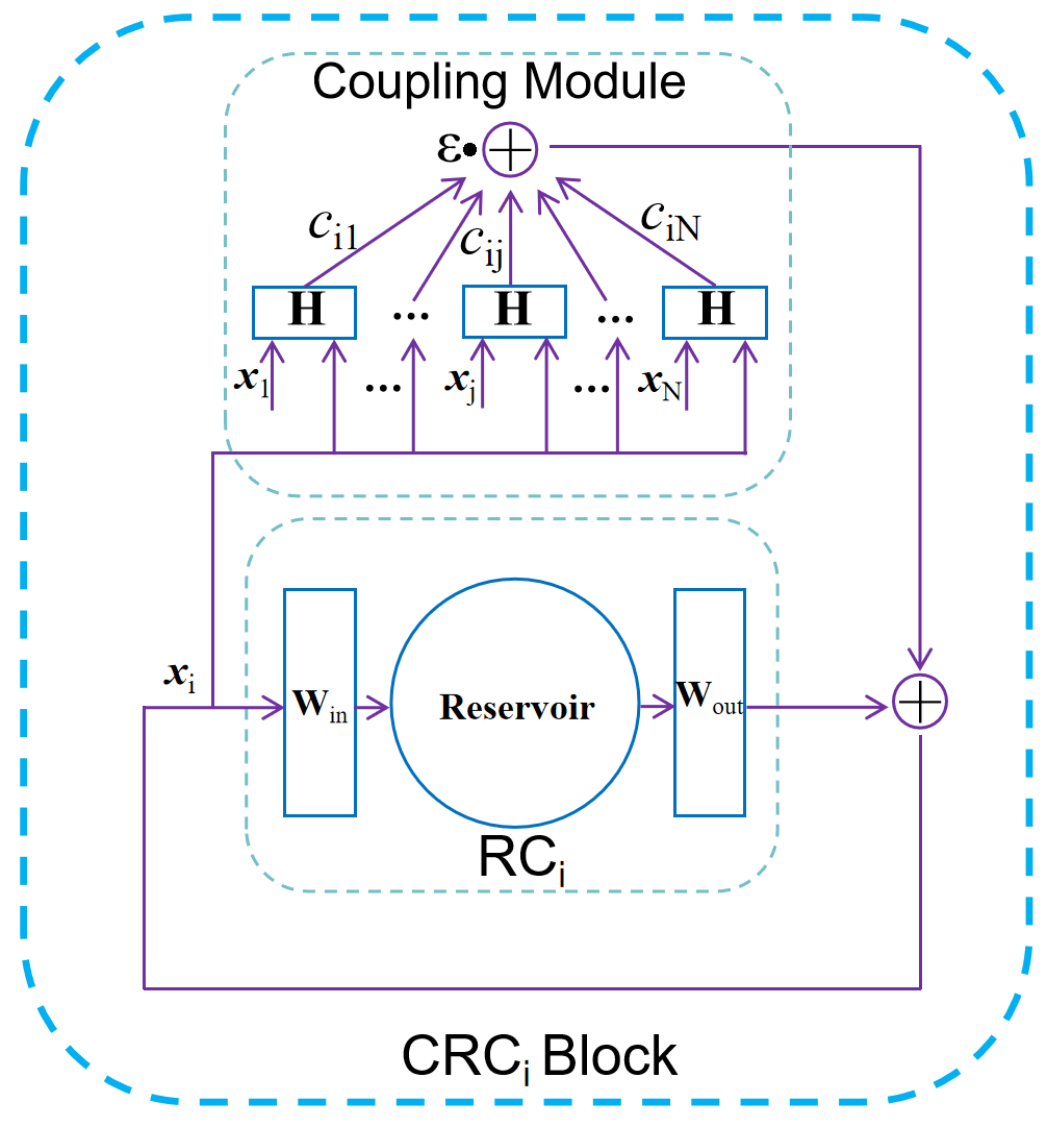}
\caption{Schematic of the coupled reservoir computing (CRC) architecture. The $CRC_{i}$ block consists of an $RC_{i}$ module for simulating the single-node dynamics and the coupling module mimicking the coupling term, whose outputs are filtered by the coupling matrix $\bm{C}$, scaled by the coupling strength $\varepsilon$, and fed into the $RC_{i}$ module.
 }
\label{fig1}
\end{center}
\end{figure}

To meet this challenge, we propose a machine learning architecture of coupled reservoir computing (CRC), which is constructed by mimicking the three constituent parts of a coupled system, namely the nodal dynamics, the coupling function, and the coupling network. More specifically, the architecture consists of $N$ single reservoir computers (RC), the coupling module, and the coupling matrix: the single RC simulates the nodal dynamics~\cite{TYHN:2019,YHBT:2024}, the coupling module incorporates the coupling information among oscillators into the evolution of a single RC and plays the same role as the coupling function, and the coupling matrix filters the outputs of the coupling module, as schematically shown in Fig.~\ref{fig1}. The single RC and the coupling module are trained on the observed time series, learning the nodal dynamics and the coupling function, respectively, whereas the coupling matrix is prescribed, as the network structure can be obtained by existing measurement and reconstruction techniques. In this design, once the three ingredients are faithfully replicated, the emergent collective behaviors, including the synchronization transitions, are expected to be reproduced spontaneously.

In this paper, we demonstrate that the CRC architecture is capable of predicting the synchronization transitions of coupled chaotic oscillators on complex networks. We show that the trained CRC reproduces not only the synchronous states, including complete synchronization, phase synchronization, cluster synchronization, and chimera states, but also the transition points as the coupling strength varies, and the predicted collective behaviors are practically identical to those obtained by direct simulations of the original system.

We consider a network of $N$ coupled chaotic oscillators~\cite{ADKM:2008,BKOV:2002}, whose dynamics is governed by
\begin{equation}\label{eq:model}
\dot{\bm{x}}_{i} = \bm{F}(\bm{x}_{i}; \bm{p}_{i}) + \varepsilon \sum_{j=1}^{N} c_{ij}\,\bm{H}(\bm{x}_{i},\bm{x}_{j}), \qquad i = 1, \ldots, N,
\end{equation}
where $i, j = 1, \ldots, N$ are the oscillator indices, $\bm{x}_{i}$ denotes the state vector of oscillator $i$, $\bm{p}_{i}$ is the parameter vector of oscillator $i$, $\bm{F}$ describes the nodal dynamics, $\bm{H}$ defines the coupling function, $\varepsilon$ is the coupling strength, and $c_{ij}$ is the coupling weight from oscillator $j$ to oscillator $i$, that is, the element in the $i$-th row and $j$-th column of the coupling matrix $\bm{C}$. The oscillators are identical when $\bm{p}_{i} = \bm{p}$ for all $i$, and have non-identical parameters when $\bm{p}_{i} \neq \bm{p}_{j}$ for $i \neq j$. The coupling matrix $\bm{C}$ is an $N \times N$ matrix that specifies the interaction structure of the network, and is prescribed throughout this paper, as the network structure can be obtained by existing measurement or reconstruction techniques. The nodal dynamics $\bm{F}$ and the coupling function $\bm{H}$ are unknown, and the task is to reproduce the collective dynamics of Eq.~\eqref{eq:model} from the observed time series.

To this end, the CRC architecture is composed of $N$ blocks, one for each oscillator. Each block consists of a single RC and a coupling module, as schematically shown in Fig.~\ref{fig1}. The single RC retains the traditional structure consisting of an input layer, a reservoir, and an output layer, and is adopted to simulate the nodal dynamics $\dot{\bm{x}}_{i} = \bm{F}(\bm{x}_{i}; \bm{p}_{i})$. The coupling module comprises $N$ parallel single-layer neural networks, each consisting of an input layer and an output layer and simulating the coupling function for one pair of oscillators. Restricting ourselves to pairwise coupling, the $j$-th neural network in the coupling module of block $i$ takes the states of oscillators $i$ and $j$ as input and outputs the coupling signal $\bm{H}(\bm{x}_{i},\bm{x}_{j})$. These signals are weighted by the $i$-th row of the coupling matrix $\bm{C}$ and scaled by the coupling strength $\varepsilon$, and the sum is incorporated into the evolution of the RC of oscillator $i$, giving rise to the coupled evolution of $N$ RCs on the complex network. Detailed descriptions of the architecture and the training scheme are provided in the Supplemental Material~\cite{ref_supp}.

\begin{figure}[tbp]
\begin{center}
\includegraphics[width=\linewidth]{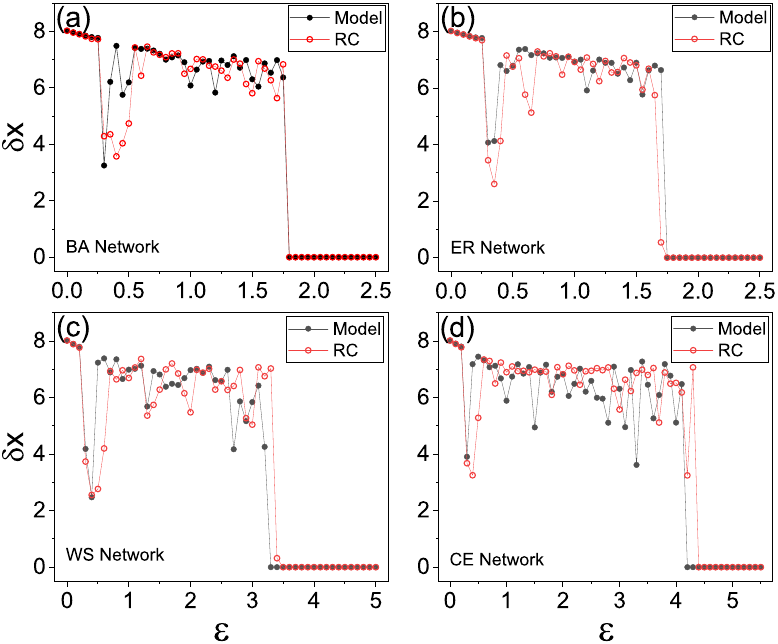}
\caption{Complete synchronization transitions of identical chaotic Lorenz oscillators coupled on complex networks. The global synchronization error $\delta x$ as a function of the coupling strength $\varepsilon$ for (a) the BA scale-free network, (b) the ER random network, (c) the WS small-world network, and (d) the C. elegans neural network. Black dots are the results obtained by direct numerical simulations, and red circles are the results predicted by the CRC.
}
\label{fig2}
\end{center}
\end{figure}

We start by predicting the complete synchronization transitions of identical chaotic oscillators coupled on complex networks. The nodal dynamics is the chaotic Lorenz oscillator, $\dot{x} = \sigma(y - x)$, $\dot{y} = x(\rho - z) - y$, and $\dot{z} = xy - \beta z$, with the parameters $\sigma = 10$, $\rho = 28$, and $\beta = 2$, by which the oscillator presents chaotic motion. The coupling function is the negative feedback diagonal coupling, $\bm{H}(\bm{x}_i, \bm{x}_j) = \bm{x}_j - \bm{x}_i$, where the components are coupled in a one-to-one manner ($x$ to $x$, $y$ to $y$, and $z$ to $z$). To examine the generality of the CRC, we consider four complex networks, namely the Barab\'{a}si-Albert (BA) scale-free network~\cite{BA:1999} ($N = 100$, $m = 5$), the Erd\H{o}s-R\'{e}nyi (ER) random network~\cite{ER:1959} ($N = 100$, $p = 0.1$), the Watts-Strogatz (WS) small-world network~\cite{WS:1998} ($N = 100$, $k = 5$, rewiring probability $p_{\rm ws} = 0.3$), and the C. elegans connectome~\cite{WSTB:1986} ($N = 279$). The coupling matrix $\bm{C}$ is the row-normalized adjacency matrix of the network, that is, $c_{ij} = a_{ij} / \sum_{j} a_{ij}$, where $a_{ij} = 1$ if oscillators $i$ and $j$ are directly connected and $a_{ij} = 0$ otherwise. In simulations, Eq.~\eqref{eq:model} is solved by the fourth-order Runge-Kutta method with the time step $\Delta t = 0.01$, and the data are collected after discarding a transient of $10^4$ time steps. To characterize the synchronization transition, we adopt the global synchronization error $\delta x = T^{-1} \sum_{t=1}^{T} [N(N-1)]^{-1} \sum_{i \neq j} |x_i(t) - x_j(t)|$, where $x_i$ is the $x$ component of oscillator $i$ and $T = 10^4$ is the length of the time window over which the average is taken. As the coupling strength $\varepsilon$ increases, $\delta x$ decreases gradually and drops close to zero at a critical coupling strength $\varepsilon_c$, signifying the onset of complete synchronization.

The training of the CRC consists of two steps, with the details provided in the Supplemental Material~\cite{ref_supp}. First, the single RC is trained on the time series of an isolated oscillator in the same way as the traditional RC, where the state vector at the previous time step is taken as the input and the state vector at the next time step is taken as the output. The trained RC is then reused as the single RC in each block. Second, the coupling module of $N$ parallel single-layer neural networks is trained on the time series of a pair of coupled oscillators at a fixed coupling strength $\varepsilon_0 = 0.35$. With the trained single RC providing the prediction of the nodal evolution, the neural network takes the states of the two oscillators as input and outputs the increment caused by the coupling, which is computed as the deviation between the actual next state and the prediction of the single RC. Once trained, the nodal dynamics $\bm{F}$ and the coupling function $\bm{H}$ are reused for all four networks, and the synchronization transitions are predicted by specifying the corresponding coupling matrix $\bm{C}$ and increasing the coupling strength $\varepsilon$.

Shown in Fig.~\ref{fig2} are the variations of $\delta x$ with respect to $\varepsilon$ predicted by the CRC (red circles) and obtained by direct numerical simulations (black dots) for the four networks. For the BA network shown in Fig.~\ref{fig2}(a), it is seen that the CRC predictions agree well with the numerical results over the entire range of $\varepsilon$. In particular, the CRC captures the weak synchronization behavior at small coupling strengths, where $\delta x$ decreases only slightly from its value in the uncoupled case, and predicts accurately the critical coupling strength at which $\delta x$ drops close to zero. Similar results are also observed for the ER, WS, and C. elegans networks in Fig.~\ref{fig2}(b-d), indicating that the trained CRC reproduces the synchronization transitions of the system on different network structures.

\begin{figure}[tbp]
\begin{center}
\includegraphics[width=\linewidth]{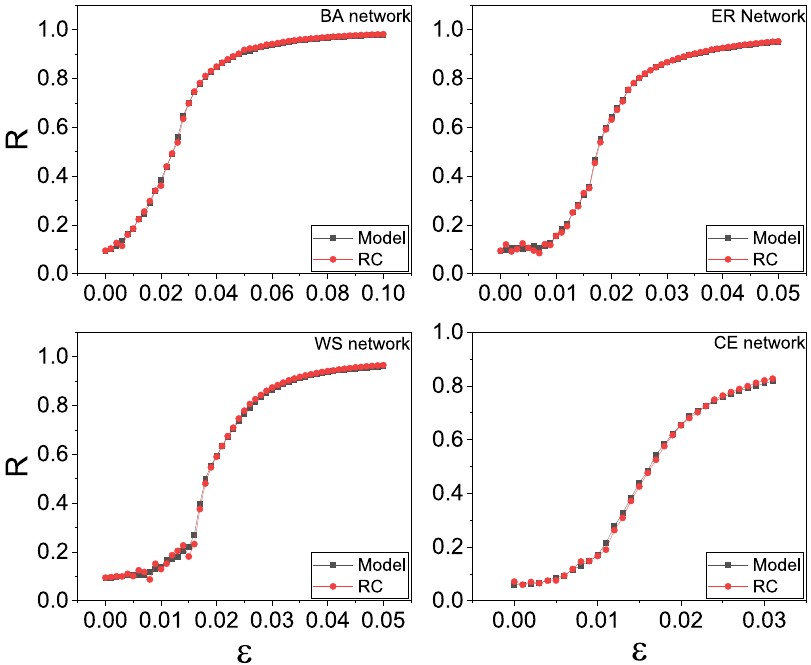}
\caption{Phase synchronization transitions of non-identical chaotic R\"{o}ssler oscillators coupled on complex networks. The synchronization order parameter $R$ as a function of the coupling strength $\varepsilon$ for (a) the BA scale-free network, (b) the ER random network, (c) the WS small-world network, and (d) the C. elegans connectome. Black squares are the results obtained by direct numerical simulations, and red circles are the results predicted by the CRC.
}
\label{fig3}
\end{center}
\end{figure}

We next consider the phase synchronization of non-identical chaotic R\"{o}ssler oscillators~\cite{RPK:1996,OHZI:2003} coupled on complex networks. The nodal dynamics is the chaotic R\"{o}ssler oscillator, $\dot{x} = -\omega_{i} y - z$, $\dot{y} = \omega_{i} x + a y$, and $\dot{z} = b + z(x - c)$, with the parameters $a = 0.15$, $b = 0.2$, and $c = 10$, by which the oscillator presents chaotic motion. The oscillators are non-identical, with the natural frequencies $\omega_i$ drawn uniformly from the range $[0.9, 1.1]$. Different from the previous case, the coupling is applied only to the $x$ components, that is, $\bm{H}(\bm{x}_i, \bm{x}_j) = (x_j - x_i, 0, 0)^{\mathrm{T}}$. Since the oscillators are non-identical, complete synchronization is no longer attainable, and we focus on the phase synchronization. Different from the previous case, the coupling matrix $\bm{C}$ is the unweighted adjacency matrix, that is, $c_{ij} = 1$ if oscillators $i$ and $j$ are directly connected and $c_{ij} = 0$ otherwise. To characterize the phase synchronization, we adopt the synchronization order parameter $R = \langle |N^{-1} \sum_{j} e^{i\theta_j}| \rangle_T$, where $\theta_j = \operatorname{atan2}(y_j, x_j) \in (-\pi, \pi]$ is the phase of oscillator $j$ obtained from its $x$ and $y$ components and $T = 10^4$ is the length of the time window over which the average is taken. As the coupling strength $\varepsilon$ increases, $R$ increases from a small value and approaches unity at a critical coupling strength $\varepsilon_c$, signifying the onset of phase synchronization.

The training procedure follows the same two steps as above. Different from the previous case, the single RC is trained in a parameter-aware manner, where the parameter is included as an additional input to the RC, and the RC is trained on the time series of isolated oscillators with different parameters, so that a single trained RC is applicable to all the oscillators despite their different parameters. The coupling module is trained in the same way as above, with details provided in the Supplemental Material~\cite{ref_supp}. Once trained, the nodal dynamics $\bm{F}_{i}$ and the coupling function $\bm{H}$ are reused for all four complex networks, and the phase synchronization transitions are predicted by specifying the corresponding coupling matrix $\bm{C}$ and increasing the coupling strength $\varepsilon$.

Shown in Fig.~\ref{fig3} are the variations of $R$ with respect to $\varepsilon$ predicted by the CRC (red circles) and obtained by direct numerical simulations (black squares) for the four complex networks. For the BA scale-free network shown in Fig.~\ref{fig3}(a), it is seen that the CRC predictions agree well with the numerical results over the entire range of $\varepsilon$. In particular, the CRC captures not only the variation of $R$ in the desynchronized and synchronized regimes but also the behavior in the vicinity of the synchronization transition point. Similar results are also observed for the ER, WS, and C. elegans connectome networks in Fig.~\ref{fig3}(b-d), indicating that the parameter-aware CRC reproduces the phase synchronization transitions of non-identical oscillators on different complex networks.

\begin{figure}[tbp]
\begin{center}
\includegraphics[width=\linewidth]{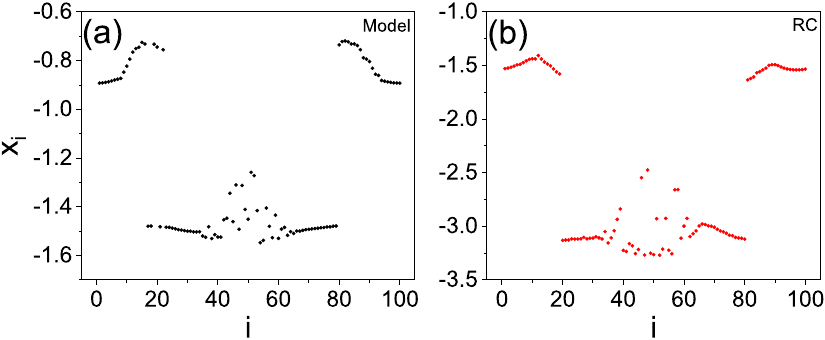}
\caption{Chimera state in a ring of identical chaotic R\"{o}ssler oscillators. (a) Snapshot of $x_i$ along the ring at a given time, obtained by direct numerical simulation (black dots). (b) The corresponding snapshot predicted by the CRC (red dots). The ring contains $N = 100$ oscillators, each coupled to its $P = 28$ nearest neighbors in each direction with coupling strength $\varepsilon = 0.06$.
}
\label{fig4}
\end{center}
\end{figure}

We finally consider the chimera state, in which synchronized and desynchronized domains coexist~\cite{OMHS:2011,ORHM:2012}. The nodal dynamics is the identical chaotic R\"{o}ssler oscillator, with $\omega_i = 1$ for all $i$ and the parameters $a = 0.42$, $b = 2.0$, and $c = 4.0$, which differ from those used above. The coupling acts on all three components, that is, $\bm{H}(\bm{x}_i, \bm{x}_j) = \bm{x}_j - \bm{x}_i$. The complex network is a ring of $N = 100$ oscillators, where each oscillator is coupled to its $P$ nearest neighbors in each direction, and the coupling matrix $\bm{C}$ is the adjacency matrix of the ring. As control parameters, we use the coupling strength $\varepsilon$ and the coupling radius $r = P/N$. Varying $P$ from 1 to $N/2$ covers the range of network topologies from local (nearest-neighbor) coupling to global (all-to-all) coupling. In what follows, we set $\varepsilon = 0.06$ and $r = 0.28$, that is, $P = 28$, with which the ring exhibits the chimera state. Following the same two-step procedure as above, the CRC is trained on the time series of an isolated oscillator and of a pair of coupled oscillators with the present parameters. To display the chimera state, we plot the snapshots of $x_i$ along the ring at a given time. Shown in Figs.~\ref{fig4}(a) and \ref{fig4}(b) are the snapshots obtained by direct numerical simulation and predicted by the CRC, respectively. It is seen that $x_i$ is scattered in the middle of the snapshot, forming an incoherent domain, whereas it varies smoothly at both ends, forming a coherent domain that is connected through the periodic boundary of the ring. The CRC reproduces the chimera pattern faithfully, including the extents of the coherent and incoherent domains. These results demonstrate that the CRC is capable of reproducing the chimera state in a ring of identical oscillators.

Compared with existing machine learning approaches for coupled systems, the CRC possesses several distinct features. First, the factorization of the architecture allows the trained modules to be reused: the same trained single RC and coupling module can be assembled with different coupling matrices to predict the collective dynamics on different complex networks, without retraining. For instance, the CRC trained once was applied to the BA, ER, WS, and C. elegans networks, and the synchronization transitions on all four networks were predicted by simply specifying their coupling matrices. Second, the training is simple and the required data are minimal: the single RC is trained on the time series of an isolated oscillator, and the coupling module on the time series of a pair of coupled oscillators, so that the amount of training data is independent of the network size, yet the trained CRC predicts the collective dynamics of large-scale networks with hundreds of oscillators. Third, the parameter-aware training of the single RC enables a single trained RC to be applied to non-identical oscillators, without training a separate RC for each oscillator. For instance, a single RC trained on isolated oscillators with a range of natural frequencies was applied to all the oscillators in the non-identical networks, despite their different frequencies. Fourth, the nodal dynamics and the coupling function are learned directly from the observed time series, without assuming their functional forms, which makes the twin applicable to systems whose governing equations are unknown. These features make the CRC particularly suitable for realizing digital twins of complex systems. A remaining issue is that the coupling matrix is prescribed in this work. This assumption, however, does not restrict the applicability of the CRC, as the coupling network is treated as a separate module that can be specified by various existing reconstruction techniques, such as compressive sensing~\cite{WLGY:2011}, Granger causality~\cite{R:2018}, and machine learning~\cite{ZZLW:2019}.

To summarize, we have proposed a machine learning architecture of coupled reservoir computing for constructing digital twins of coupled oscillator systems, and have demonstrated its capability in predicting synchronization transitions. Once trained on the observed time series, the CRC reproduces the complete synchronization transitions of identical oscillators on different complex networks, the phase synchronization transitions of non-identical oscillators, and the chimera state in a ring, with the predicted behaviors practically identical to those obtained by direct numerical simulations. In addition, the CRC predicts synchronization patterns, e.g., cluster synchronization, as demonstrated in the Supplemental Material~\cite{ref_supp}. These results demonstrate that the CRC reaches the third and fourth levels of fidelity for a coupled system, which have so far been beyond the reach of existing machine learning approaches for large-scale coupled systems. It is our hope that the CRC architecture will serve as a useful tool for the study of collective dynamics in complex systems.

\section*{Acknowledgements}
This work was supported by the National Natural Science Foundation of China (NNSFC) under Grant Nos.~12105165 and 12275165. H.F. was also supported by the Natural Science Basic Research Plan in Shaanxi Province of China under Grant No. 2026JC-YBQN-0016.

\bibliographystyle{apsrev4-1}
\bibliography{ML_NLD}

\end{document}